\documentclass[a4paper,10pt]{article}

\usepackage{amsfonts}
\usepackage{graphicx}
\usepackage{amsmath}
\usepackage{amssymb}
\usepackage{amsthm}
\usepackage{newlfont}
\usepackage{stmaryrd}
\usepackage{mathrsfs}
\usepackage{euscript}

\begin{document}
%opening
\title{The modified Sutherland--Einstein relation\\ for diffusive nonequilibria}
\author{Marco Baiesi$^1$, Christian Maes$^2$, Bram Wynants$^3$\\ \\
$^1$ Dipartimento di Fisica, Universit\`a di Padova,\\ Via Marzolo 8, 35131 Padova, Italy\\
$^2$ Instituut voor Theoretische Fysica, K.U.Leuven,\\ 3001 Leuven, Belgium \\
$^3$ Institut de Physique Th\'eorique CEA-Saclay,\\ F-91191 Gif-sur-Yvette Cedex, France \\
\textit{bram.wynants@cea.fr}
}
\date{}
\maketitle

\begin{abstract}
There remains a useful relation between diffusion and mobility for a Langevin particle in a periodic medium subject to nonconservative forces.  The usual fluctuation-dissipation relation easily gets modified and the mobility matrix is no longer proportional to the diffusion matrix, with a correction term depending explicitly on the (nonequilibrium)
forces. We discuss this correction by considering various simple examples and we visualize the various dependencies on the applied forcing and on the time by means of simulations. For example, in all cases the diffusion depends on the external forcing more strongly than does the mobility. We also give an explicit decomposition of the symmetrized mobility matrix as the difference between two positive matrices, one involving the diffusion matrix, the other force--force correlations.
\end{abstract}

\section{Introduction}
 The relation between internal fluctuations in a system on the one hand and its susceptibility under an applied external field on the other hand is of great practical and theoretical interest.  The most immediate example concerns the relation between diffusion and mobility and remains important in the development of nonequilibrium statistical mechanics. The first studies go back to the works of \cite{sut1,sut2} and of \cite{ein}, (see also \cite{mck}).  They are early examples of the fluctuation--dissipation theorem, itself being a cornerstone of linear response theory for equilibrium systems, see \cite{kub} and \cite{spo}.\\
   When moving to nonequilibrium systems, the fluctuation--dissipation relation (FDR) is typically violated and there is no {\it a priori} reason why the system's mobility can be simply obtained from its diffusivity.  All the same, mobility remains crucially important for the discussion of transport properties under nonequilibrium conditions \cite{gea,spo}. That is why we want to get explicit information about the corrections to the FDR, a subject that has been considered from many different sides.  We mention few.  One approach has been to associate an effective temperature to the nonequilibrium system, for example in ageing regimes, as e.g. in \cite{ag} and \cite{oi}.  Another approach is to modify the FDR inserting the nonequilibrium stationary density, or going to a ``moving'' frame of reference using the probability current or local velocity \cite{bea3}, somewhat in the spirit of the approach of \cite{aga}, see also \cite{ht,kf1,kf2,ss2}. Specific models of interacting particles have for example been treated, also in mathematically rigorous ways, by \cite{he,ko,lr,lou}. More recent experimental work includes \cite{gea,mea,oi}.\\
     Here we concentrate on the nonequilibrium regime where the Sutherland--Einstein relation gets modified because of nonconservative forces, starting from the inertial regime.  As will become clear, not much effort is needed for finding examples where
these corrections are visible and detectable. Recent work by \cite{kf1,kf2} is similar to ours but in a different setting. The present paper gives the corrections following recent work by \cite{bea1,bea2}, which enables us to zoom in on particular dependencies, such as on environment and driving parameters.\\
The main results of the present analysis are the explicit relations (\ref{genresult2}) and (\ref{symr}) between the mobility and the diffusion constant, and their numerical exploration.
 In particular, we typically find that for the models treated here the diffusion constant depends on the external (nonequilibrium) forcing in a stronger way than the mobility. The mobility is bounded from above by the diffusion, as we derive in an exact bound (\ref{bou}). When the forcing only depends on the position, then the correction to the Sutherland--Einstein relation is second order. \\

 The plan of the paper is as follows.  Section \ref{modein} contains the general relation between mobility and diffusion matrix for Langevin dynamics in a magnetic field plus other general external forces. The correction to the Sutherland--Einstein relation is in terms of a correlation function between the particle velocity and the applied forces. As simplest and standard illustration Section \ref{clas} reminds us of the theoretical framework of the Sutherland--Einstein relation.  It is the famous proportionality of the mobility (in linear regime) with the diffusion constant (of the original system). We see from these textbook examples that our approach is suitable. Then restarts the nonequilibrium analysis.  Section \ref{uni} treats nonequilibrium fluids with uniform temperature.  The nonequilibrium condition is imposed by stirring the fluid (rotational or nonconservative forces).  Here our motivation is primarily to illustrate the existing exact formul{\ae} and to explore more the role and the influence of the various parameters.  All involved observables can be measured under the nonequilibrium averaging, and vice versa, the formul{\ae} in principle enable to learn about unknown driving forces and parameters from the correction to the Sutherland--Einstein relation.  Section \ref{app2} explores the symmetrized
 mobility; the diagonal elements allow some general bound.  Finally, the Appendix contains some more technical points dealing with analysis and numerics.\\

\section{A general mobility-diffusion relation}\label{modein}

Consider a particle of mass $m$, which diffuses in a heat bath (e.g. some fluid) in $\mathbb{R}^3$ according to the Langevin dynamics
\begin{eqnarray}
\dot{\vec{r}}_t = \frac{\vec{r}_{t+dt} -\vec{r}_t}{dt} &=& \vec{v}_t \label{general}\\
 m\dot{\vec{v}}_t  = m\frac{\vec{v}_{t+dt} -\vec{v}_t}{dt} &=& \vec{F}(\vec{r}_t,\vec{v}_t)-\gamma m \vec{v}_t + \sqrt{2m\gamma T}\,\vec{\xi}_t \nonumber
\end{eqnarray}
Here $\vec{r}_t$ is the position of the particle at time $t$, and $\vec{v}_t$ is its velocity. The particle is passive and undergoes the influence of a heat bath in thermal equilibrium at temperature $T$.  In the usual weak coupling regime the particle suffers friction with coefficient $\gamma$ and random collisions, here represented by the vector $\vec{\xi}_t$ of standard Gaussian white noises. This means that each of the components has a Gaussian distribution,
with mean zero $\left<\xi_{t,i}\right> = 0$ and covariance $\left<\xi_{t,i}\xi_{s,j}\right> = \delta_{i,j}\delta(t-s)$.
Throughout this text we set Boltzmann's constant to one.
Furthermore there is an external force $\vec{F}$ working on the particle. Throughout this text, we restrict ourselves to forces that depend periodically on the position $\vec{r}$.
We make this restriction to have a form of translation invariance in the system. In other words, we consider time-scales over which there is no confining potential.  The force $\vec{F}$ can depend on the velocity as for a magnetic field when the particle is charged.\\

When given an initial (probability) density $\mu(\vec{r},\vec{v})$ of independent such particles at time zero, it changes in time according to the Fokker-Planck equation
\begin{eqnarray}\label{fokplagen}
 \frac{\partial \mu_t}{\partial t} = -\vec{v}\cdot\vec{\nabla}_r\mu_t-\vec{\nabla}_v\cdot \Big[\Big(\frac{\vec{F}-m\gamma  \vec{v}}{m}\Big)\mu_t - \frac{\gamma T}{m}\vec{\nabla}_v \mu_t\Big]
\end{eqnarray}
There are two important quantities that characterize the transport behavior of such a system.
The first quantity is the diffusion (matrix) function $D(t)$, which is defined as
\[
 D_{ij}(t) = \frac{1}{2t}\Big<(\vec{r}_t-\vec{r}_0)_i;(\vec{r}_t-\vec{r}_0)_j\Big>
  \]
The subscripts denote the components of the corresponding vectors, and the right-hand side is a truncated
correlation function: for observables $A$ and $B$
\begin{equation}\label{trunc}
 \Big<A;B\Big>=\Big<AB\Big> -\Big<A\Big>\Big< B\Big>
 \end{equation}
For diffusive systems, i.e., for the systems described above, this diffusion function
is expected to have a large time limit, called the diffusion matrix
\[ D_{ij} = \lim_{t\to\infty}D_{ij}(t) \]
In words, the (co)variance of the displacement of the particle is linear with time (for large times $t\gg 1/\gamma$)
with slope given by the diffusion constant. Some analysis is found in Appendix \ref{app1}.
Secondly, there is the mobility (matrix) function $M(t)$, defined as follows:
we add to the dynamics in (\ref{general}) a constant (but small) force $\vec{f}$, replacing $\vec{F}(\vec{r},\vec{v})
\rightarrow \vec{F}(\vec{r},\vec{v}) + \vec{f}$. The mobility
then measures the change in the expected displacement of the particle:
\[ M_{ij}(t) = \frac{1}{t}\left.\frac{\partial}{\partial f_j}\Big<({\vec r}_t-{\vec r}_0)_i\Big>^{f}\right|_{\vec{f}=0} \]
where the superscript $f$ wants the average to be taken in the dynamics with the extra force $\vec{f}$.
Again, this function is supposed to have a large-time limit, which is called the mobility:
\[ M_{ij} = \lim_{t\to\infty}M_{ij}(t) \]
i.e., the mobility is the linear change in the stationary velocity by the addition of a
small constant force.

In the special case of detailed balance dynamics, these two quantities are related by the Sutherland--Einstein relation,
\begin{equation}\label{einstein} M_{ij} = \frac{1}{T}D_{ij} \end{equation}
which is an instance of the more general fluctuation-dissipation theorem.\\
When the system is not in equilibrium, mobility and diffusion constants are no longer proportional.
Additional terms show up: for the dynamics defined in (\ref{general}) we get
\begin{eqnarray}
 && M_{ij}(t) = \frac{1}{T}D_{ij}(t) +\frac{1}{2\gamma mT t}\Big<({\vec r}_t - {\vec r}_0)_i;{\vec \Psi}_j\Big> \label{genresult1}
\end{eqnarray}
where the vector $\vec{\Psi}$ is explicitly given by
\begin{equation}\label{psi}
\vec{\Psi} = m(\vec{v}_t - \vec{v}_0) -\int_0^t ds\vec{F}(\vec{r}_s,\vec{v}_s)
\end{equation}
The relation between the mobility and diffusion functions is thus modified with respect to
the large time limits in equilibrium systems by
the addition of an extra term. This term is the (truncated) correlation function between
the displacement and the functional $\vec{\Psi}$. This functional is explicitly expressed
in terms of the velocity of the particle of interest and the forces that act on it.
In the limit of large times, the relation (\ref{genresult1}) simplifies somewhat: in this limit we get
\begin{eqnarray}
 && M_{ij} = \frac{1}{T}D_{ij} + \lim_{t\to\infty}\frac{1}{2\gamma mT t}\Big<({\vec r}_t - {\vec r}_0)_i;{\vec \Phi}_j\Big>\label{genresult2}
\end{eqnarray}
where the vector $\vec{\Phi}$ is explicitly given by
\begin{equation}
\vec{\Phi} =  -\int_0^t ds\,\vec{F}(\vec{r}_s,\vec{v}_s)\label{phi}
\end{equation}
Indeed, the term with the correlation between the displacement and the change of velocity does
not contribute in the large-time limit:
\begin{equation}\label{corrnul}
\lim_{t\to\infty}\frac{1}{t}\Big<({\vec r}_t - {\vec r}_0)_i;(\vec{v}_t - \vec{v}_0)_j\Big> = 0
\end{equation}
For an argument see Appendix \ref{app1}.  In Section \ref{app2} we further rewrite (\ref{genresult2}) to obtain
\begin{equation}\label{symr}
\frac{M_{ij}+M_{ji}}{2} = \frac{D_{ij}}{2T} + \frac{\delta_{i,j}}{2\gamma m} - \lim_{t\to \infty}\frac{1}{4\gamma^2 m^2T t}\Big< {\vec \Phi}_i;{\vec \Phi}_j \Big>
\end{equation}
where each term represents a symmetric matrix.  Under conservative forces the second and the third term sum to give the first term on the right-hand side.  Otherwise, in diffusive nonequilibrium the correction term to the Sutherland--Einstein relation is nonzero with the symmetrized mobility matrix as the explicit difference between a diffusion--related matrix and the force-force covariance matrix, as claimed in the last line of the abstract.
A further bound on the symmetrized mobility is added in Section \ref{app2}.\\

Our modified Sutherland--Einstein relation has the advantage of being explicit in the dynamical variables.  Still the formula is sufficiently complicated and some aspects of it require careful examination.  We can rewrite formula (\ref{genresult2}) into
\begin{eqnarray}
 && M_{ij} = \frac{1}{T}D_{ij} - \lim_{t\to\infty}\frac{1}{2\gamma mT }\,\int_0^t ds\,\Big<\frac{({\vec r}_t - {\vec r}_0)_i}{t};F_j(\vec{r}_s,\vec{v}_s)\Big>\label{genresult5}
\end{eqnarray}
As we see, the correction to the equilibrium mobility--diffusion relation is measured by a space--time correlation between applied forcing and displacement. One simplification is to look close-to-equilibrium.  There we see by time-reversal symmetry applied to the reference equilibrium that the correction is only quadratic in the applied forcing when $\vec{F}$ only depends on $\vec{r}$. In other words, the deviations with respect to the Sutherland-Einstein relation are then second order.
Going further from equilibrium also the inverse question becomes interesting, to characterize the nonequiilibrium forcing that produces a given experimentally determined mobility-diffusion relation.  We hope to see in future work that formul{\ae} like \eqref{genresult5} are also useful to learn about the driving conditions from measurements of both mobility and diffusion. Numerically, of course the right-hand side of \eqref{genresult5} is easier to determine than the left-hand side. \\

Usually, as also in what follows, one considers the diffusion and the mobility of a single particle.  It is however relevant and also possible to include interactions with other particles.  The equations (\ref{general}) must be changed for the force to include this dependence on the state of the other particles, but the main results remain unaffected.  After all, one can simply redo all calculations in larger dimensions ($3N$ for $N$ particles) and consider independent white noises acting on all 3N-components.  Furthermore, the force $F$ can be time-dependent and formulae (\ref{genresult2}) -- (\ref{symr}) remain the same but where $F$ has further time-dependencies, be it explicit or be it because of the time-dependent state of other particles.\\

The explicit relations above have been derived in \cite{bea2} following a path-integral formalism. In the present paper we
provide a number of examples to discuss this relation between mobility and diffusion out of equilibrium.
For each of these examples, simulations have been done to help visualization, and to inform us about the relative importance of the various parameters and terms in (\ref{genresult2}). On the other hand, the simulations raise new questions, not all of which we are able to answer. The Appendix adds more theoretical considerations to the relation between diffusion and mobility, and gives information about the simulation method.

\section{Classical equilibrium theory}\label{clas}

\noindent\textbf{Pure diffusion}\\
The simplest form of the Langevin equation is the case where all the external forces are set to zero
\begin{eqnarray}\label{diffusion}
 m\dot{\vec{v}}_t &=& -\gamma m \vec{v}_t + \sqrt{2m\gamma T}\,\vec{\xi}_t
\end{eqnarray}
The mobility and diffusion can be computed explicitly, by first integrating the Langevin equation
(adding a force $\vec{f}$),
\[ \vec{v}_t = \vec{v}_0e^{-\gamma t} +\frac{\vec{f}}{\gamma m}\left[1-e^{-\gamma t}\right] + \int_0^tdse^{-\gamma(t-s)}\vec{\xi}_s \]
and then using the properties of the Gaussian white noise. Because $\left<\vec{\xi}_t\right> = 0$, we immediately get the mobility by again integrating over time,
\begin{eqnarray*}
 M_{ij}(t) &=& \delta_{ij}\Big[\frac{1}{\gamma m} - \frac{1-e^{-\gamma t}}{\gamma^2 mt}\Big]\\
\end{eqnarray*}
As $\left<\xi_{i,t}\xi_{j,s}\right> = \delta_{i,j}\delta(t-s)$ the velocity-velocity correlations equal (for $\vec{f}=0$)
\[ \Big<v_{i,t}v_{j,s}\Big> = \left[ \left<v_{0,i}v_{0,j}\right> - \frac{\delta_{i,j}T}{m} \right]e^{-\gamma(t+s)} +  \frac{\delta_{i,j}T}{m}e^{-\gamma |t-s|}\]
Simply integrating over time then gives the diffusion,
\begin{eqnarray*}
 D_{ij}(t) &=& \frac{1}{2\gamma^2t}\Big[\Big<v_{0,i};v_{0,j}\Big>-\frac{T\delta_{ij}}{m}\Big]\Big[1-e^{-\gamma t}\Big]^2 +\delta_{ij}T\Big[\frac{1}{\gamma m} - \frac{1-e^{-\gamma t}}{\gamma^2 mt}\Big]
 \end{eqnarray*}
Taking the limit of large times, we see that
\[ M_{ij} = \lim_{t\to\infty}M_{ij}(t) =  \frac{\delta_{ij}}{\gamma m} = \lim_{t\to\infty} \frac{D_{ij}(t)}{T} = \frac{D_{ij}}{T}\]

\begin{figure}[!h]
\centering
\includegraphics[width=0.45\textwidth]{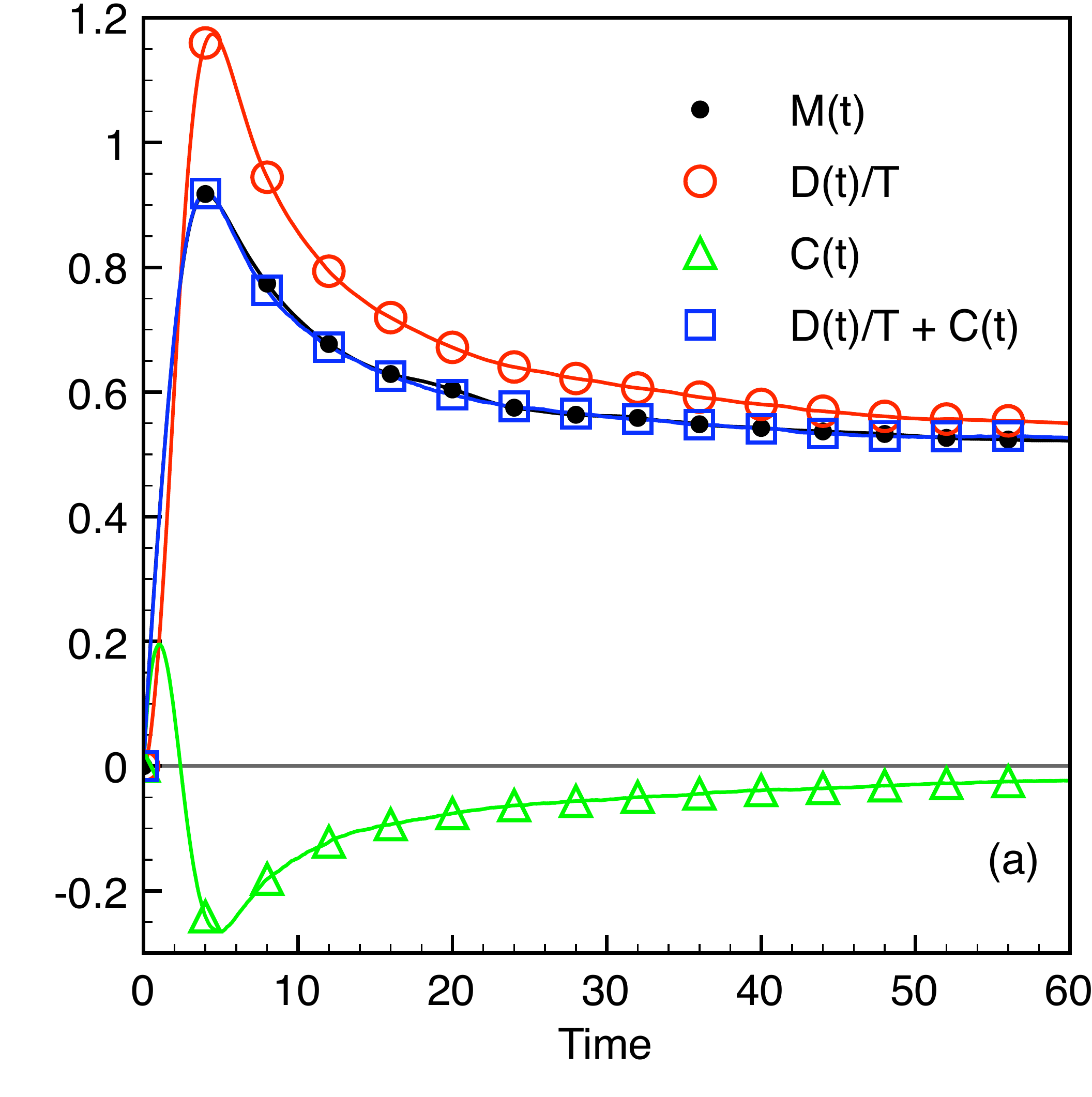}
\includegraphics[width=0.45\textwidth]{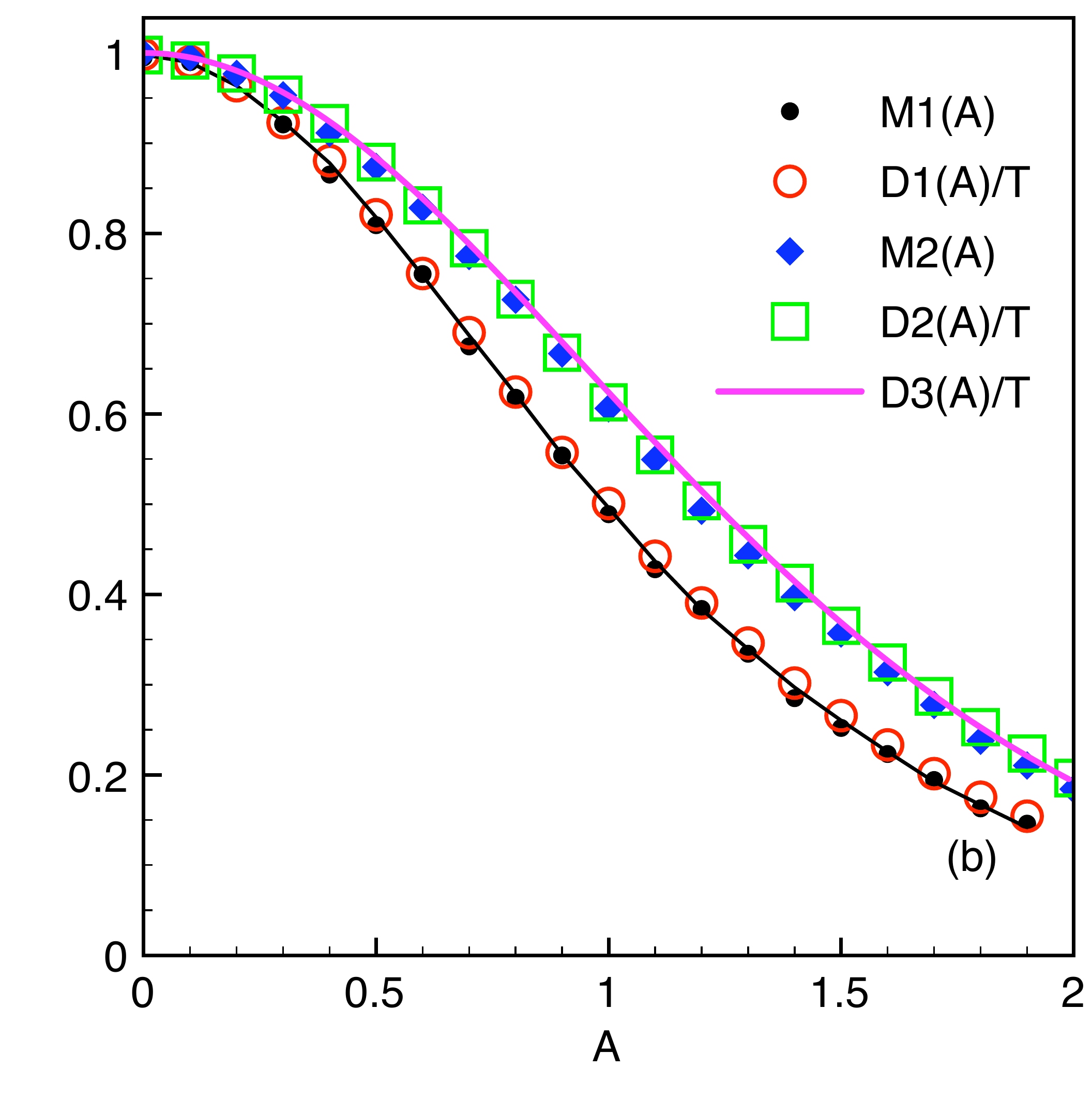}
\caption{Diffusion in potential $U(x) = A\,\cos x$.
 In (a) $A=1$ and the evolution through time is shown. Here $T=\gamma = m = 1$.
$C(t)$ here represents the last two terms on the right-hand side of (\ref{eqpot}).
In (b) the long time limits of the mobility and diffusion are shown for values of $A$ ranging from $0$ to $2$.
$M1$ and $D1$ are the mobility and diffusion for the case $T=\gamma = m = 1$, while $M2$ and $D2$ correspond to $T =1$, $\gamma = 10$, $m=0.1$. Finally $D3$ is the diffusion that is explicitly calculated by (\ref{fesgal}). (Online version in color.)}
\label{fig:diffpot}
\end{figure}

\noindent\textbf{A periodic potential}\\
When a force is added that derives from a
periodic potential $U$,
\begin{eqnarray}\label{diffpot}
\dot{\vec{r}}_t &=& \vec{v}_t\nonumber\\
 m\dot{\vec{v}}_t &=& -\nabla_r U(\vec{r}_t)-\gamma m \vec{v}_t + \sqrt{2m\gamma T}\,\vec{\xi}_t
\end{eqnarray}
then, the general expression (\ref{genresult1}) gives
\begin{eqnarray}\nonumber
 M_{ij}(t) &=& \frac{1}{T}D_{ij}(t) + \frac{1}{2\gamma T t}\Big<{\vec r}_t-{\vec r}_0)_i;({\vec v}_t-{\vec v}_0)_j\Big>\\
&& - \frac{1}{2\gamma m T t}\int_0^t ds\Big<({\vec r}_t-{\vec r}_0)_i;\nabla_{r_j} U(\vec{r})\Big>\label{eqpot}
\end{eqnarray}
The second term on the right-hand side tends to zero for large times, as said before, as does the third term:
\begin{equation} \label{corrpotzero}
\lim_{t\to\infty}\frac{1}{t}\int_0^t ds\Big<({\vec r}_t-{\vec r}_0)_i;\nabla_{r_j} U(\vec{r})\Big> =0
\end{equation}
yielding the standard $M_{ij} = D_{ij}/T$; see also Appendix \ref{app1}.
Of course the mobility and diffusion
are no longer equal to $\delta_{ij}/(\gamma m)$. We get a summary in Fig.~\ref{fig:diffpot}.
Observe for example that the mobility decreases with the amplitude of the conservative force, which is logical, since the particle
needs to escape potential wells to have a non-zero velocity.\\

In fact, in \cite{fg} an explicit expression was derived for the diffusion for overdamped Langevin
equations with periodic potentials. The overdamped Langevin equation is what one gets when the friction is high
and the mass of the particle is small, so that inertial effects can be ignored. Formally, one lets $\gamma\to\infty$
and $m\to 0$, while keeping $\chi = \frac{1}{m\gamma}$ constant and finite. The result is that one can put $m\dot{\vec{v}}_t$ to zero in (\ref{diffpot}). The resulting Langevin equation can then be written solely in terms of the position:
\[ \dot{\vec{r}}_t = -\chi\nabla_r U(\vec{r}_t)+ \sqrt{2\chi T}\,\vec{\xi}_t \]
The formula of \cite{fg} for one-dimensional diffusion is given by
\begin{equation}\label{fesgal} D = \frac{\chi T R^2}{\int dx \,e^{U(x)/T}\int dx\, e^{-U(x)/T}} \end{equation}
where $R$ is the period of the potential, and the integrals in the denominator are over one period.  We give a new and short proof of
\eqref{fesgal} in Appendix \ref{mo1}.
In Fig.~\ref{fig:diffpot} we have plotted this explicit expression, for a potential $U = A\cos(x)$
for various values of $A$. This curve corresponds nicely with the mobility and diffusion we got from simulations,
where we took $\gamma = 10$ and $m = 0.1$.\\

\begin{figure}[!h]
\centering
\includegraphics[width=0.45\textwidth]{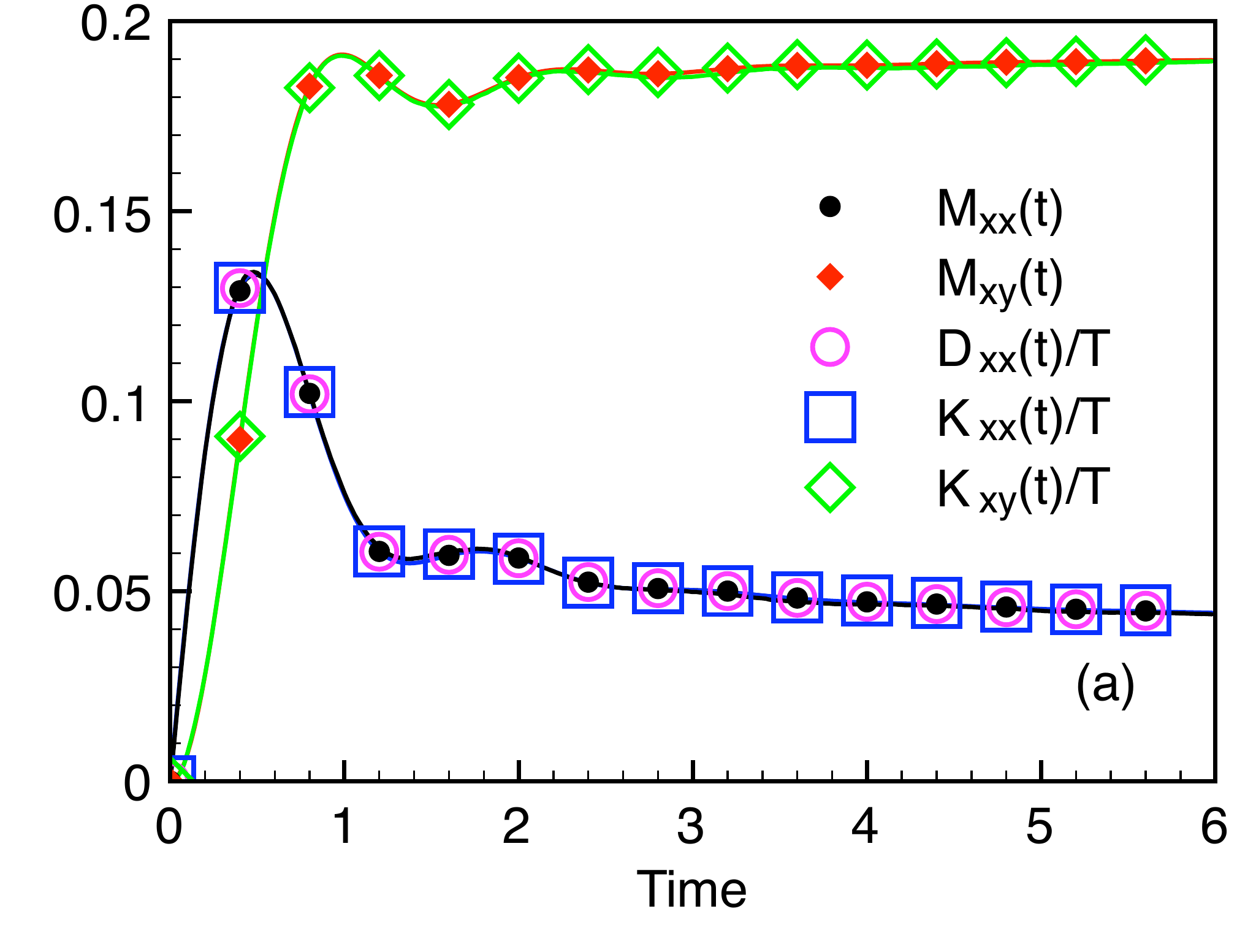}
\includegraphics[width=0.45\textwidth]{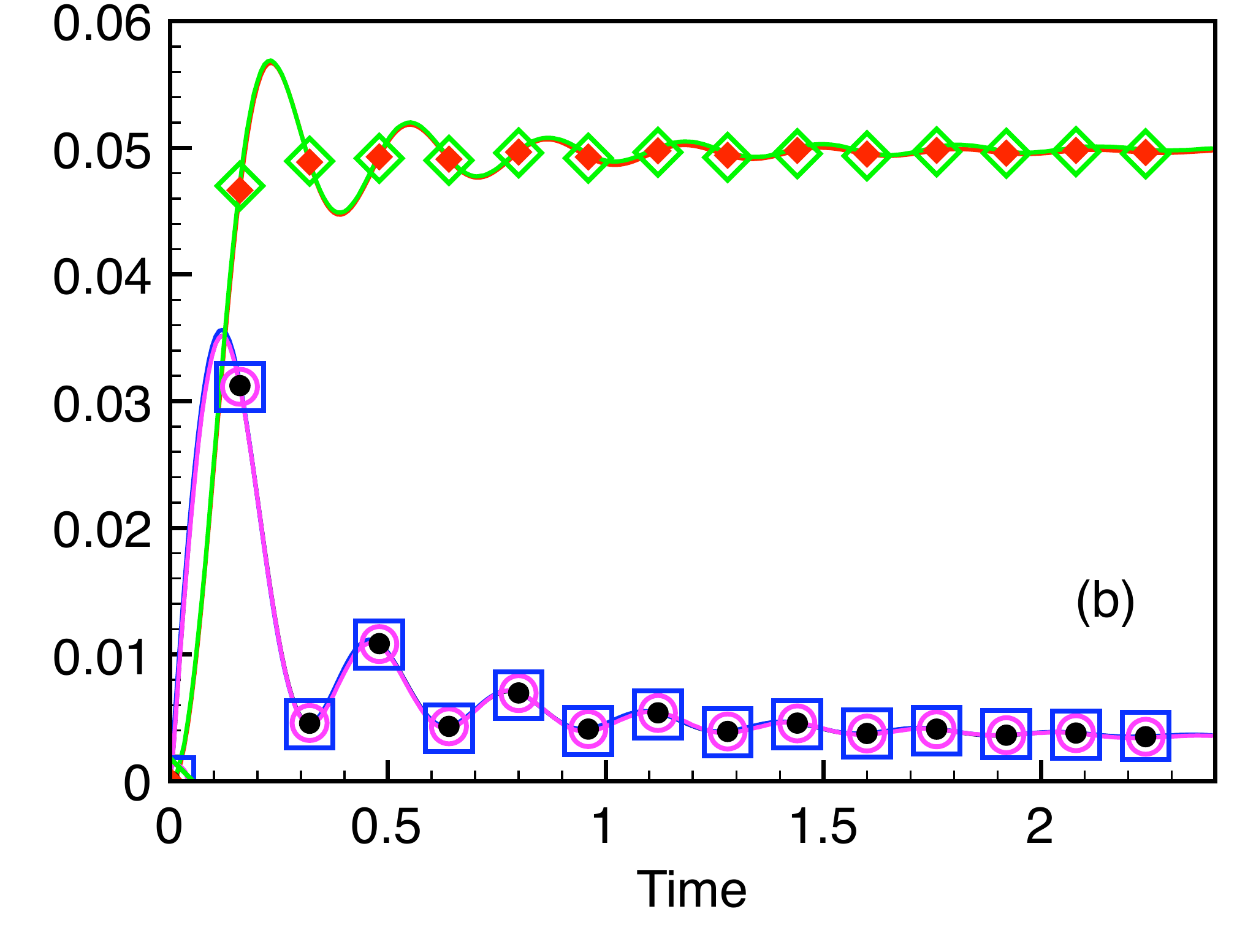}
\caption{Diffusion with a magnetic field in the $z$-direction.
Here $T=\gamma = m = 1$, while $B=5$ for Figure (a) and $B=20$ for Figure (b). (Online version in color.)}
\label{fig:diffmag}
\end{figure}

\newpage
\noindent\textbf{A magnetic field}\\
A first choice of a velocity dependent force $\vec{F}$ in \eqref{general} could be a magnetic field.  If we add to (\ref{diffusion}) a magnetic field, say a homogeneous magnetic field in the $z$-direction, $ {\vec r}_t= (x_t,y_t,z_t), \vec{B} = (0,0,B)$, with unit electric charge,
\begin{eqnarray}
 m\dot{v}_{x,t} &=&  Bv_{y,t}-\gamma m v_{x,t} + \sqrt{2m\gamma T}\,\xi_{x,t}\nonumber \\
 m\dot{v}_{y,t} &=&  - Bv_{x,t}-\gamma m v_{y,t} + \sqrt{2m\gamma T}\,\xi_{y,t}\label{langmag}
\end{eqnarray}
then, (\ref{genresult2}) immediately gives
\begin{eqnarray*}
M_{xx}=\frac{D_{xx}}{T}-\frac{BD_{xy}}{\gamma m T}, \ \ \ \ \ M_{xy}=\frac{D_{xy}}{T} + \frac{BD_{xx}}{\gamma m T}\\
M_{yx}=\frac{D_{yx}}{T}-\frac{BD_{yy}}{\gamma m T}, \ \ \ \ \ M_{yy}=\frac{D_{yy}}{T} + \frac{BD_{yx}}{\gamma m T}
\end{eqnarray*}
Further explicit calculations show that
\[ D_{xx} = D_{yy} = \frac{\gamma m T}{\gamma^2 m^2 + B^2}, \ \ \ \ \ D_{xy} = D_{yx} = 0 \]
The diagonal elements of $M$ and $D$ are proportional, but not so for the off-diagonal elements.
Therefore, adding a magnetic field to an otherwise equilibrium dynamics breaks the equilibrium Einstein relation.
This may be counterintuitive, because adding a magnetic field leaves the equilibrium Boltzmann distribution intact.
 In Fig.~\ref{fig:diffmag} the results of simulations are shown.
We took the initial position equal to zero, and the velocity is Maxwellian $\rho(\vec{v})\propto\exp(-mv^2 / 2T)$, which is the stationary velocity distribution. The $K(t)$ show the right-hand side of (\ref{genresult1}).
As one can see, they coincide nicely with the mobility (left-hand side of (\ref{genresult1})). Note that the oscillations of the mobility and diffusion have a period $2\pi m / (B T)$.

\newpage

\section{Nonconservative forces}\label{uni}

Recall that we work in unbounded spaces with periodic forces.  In that sense we consider diffusion on the torus.
A conservative force is then a force for which every contour integral that starts and ends in equivalent points equals zero. Equivalent points have the
same coordinates modulo the period of the force. If not, we call the force nonconservative.
Practically, we distinguish two kinds of nonconservative forces.
First of all there are forces that cannot be derived from a potential (or equivalently, that
have a nonzero curl) in the full space. Secondly, there are forces that can be derived from a potential in the full space but not from a potential on the torus.
We give examples of both cases in this section.

\subsection{Nonperiodic potential}

\begin{figure}[!h]
\centering
\includegraphics[width=0.465\textwidth]{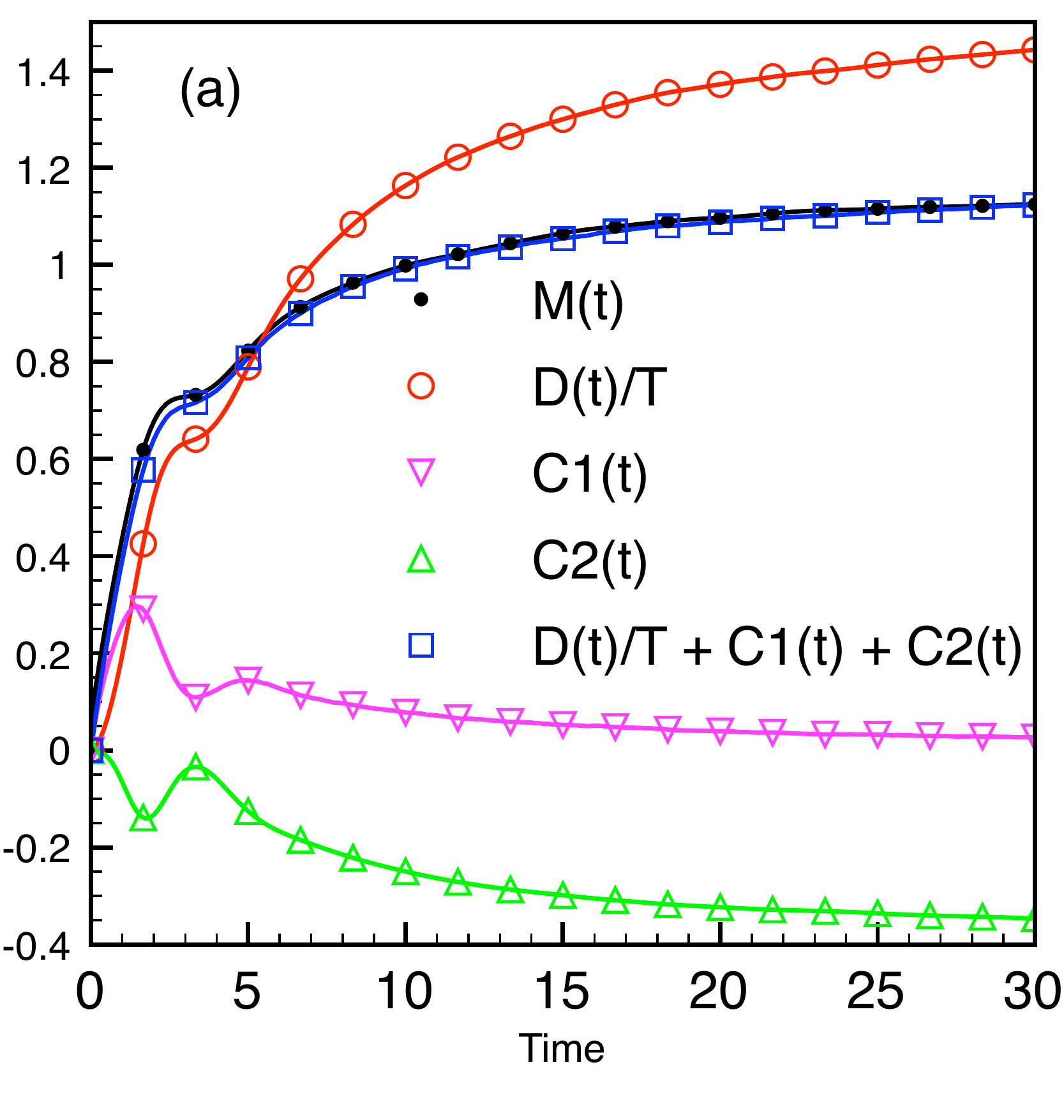}
\includegraphics[width=0.435\textwidth]{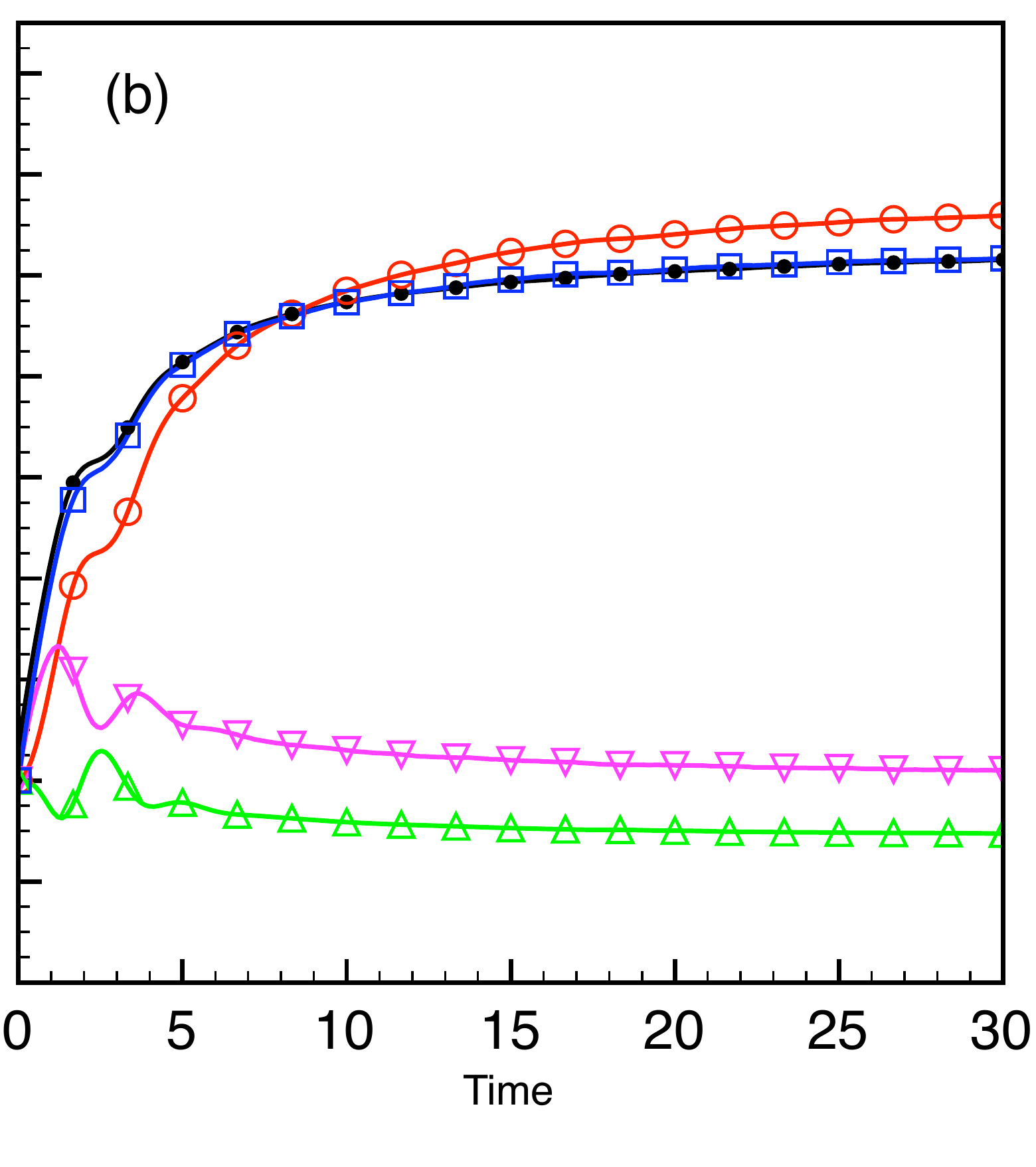}
\caption{Simulations of a diffusion with a potential $-Fx + \cos x$.
Here $T=\gamma = m = 1$. In (a) $F=1.5$ and in (b) $F=2.5$. (Online version in color.)}
\label{fig:withforce}
\end{figure}

Consider again a particle diffusing on an infinite line, with potential $U(x) = \cos x$. To this we add
a constant force $F$:
\begin{eqnarray*}
\dot{x}_t &=& v_t\\
 m\dot{v}_t &=& F + \sin x_t -\gamma m v_t + \sqrt{2m\gamma T}\,\xi_t
\end{eqnarray*}
The force $F + \sin x$  can be derived from the potential $-Fx+\cos x$,
but this potential is not periodic. The relation (\ref{genresult1}) becomes here
\[ M(t) = \frac{D(t)}{T} + \underbrace{\frac{1}{2\gamma Tt}\Big<(v_t-v_0);(x_t-x_0)\Big>}_{C1(t)}
+ \underbrace{\frac{1}{2\gamma mTt}\int_0^tds\Big<\frac{dU}{dx}(x_s);(x_t-x_0)\Big>}_{C2(t)} \]
Note that $\Big<F;(x_t-x_0)\Big> = 0$ because $F$ is a constant. We have simulated this system
for various values of the force $F$. In Fig.~\ref{fig:withforce} the mobility, diffusion, the correlation
of the displacement with the change in velocity (C1) and the correlation of the displacement with the integrated force
(C2) are shown for $F=1.5$ (a), $F=2.5$ (b). In both cases the initial conditions are $x_0=v_0=0$.
Note that all quantities converge to a constant value. This is because the diffusion and other correlations
are defined as truncated correlation functions (see (\ref{trunc})), so that the diverging parts of the correlations are subtracted.

\begin{figure}[!h]
\centering
\includegraphics[width=0.45\textwidth]{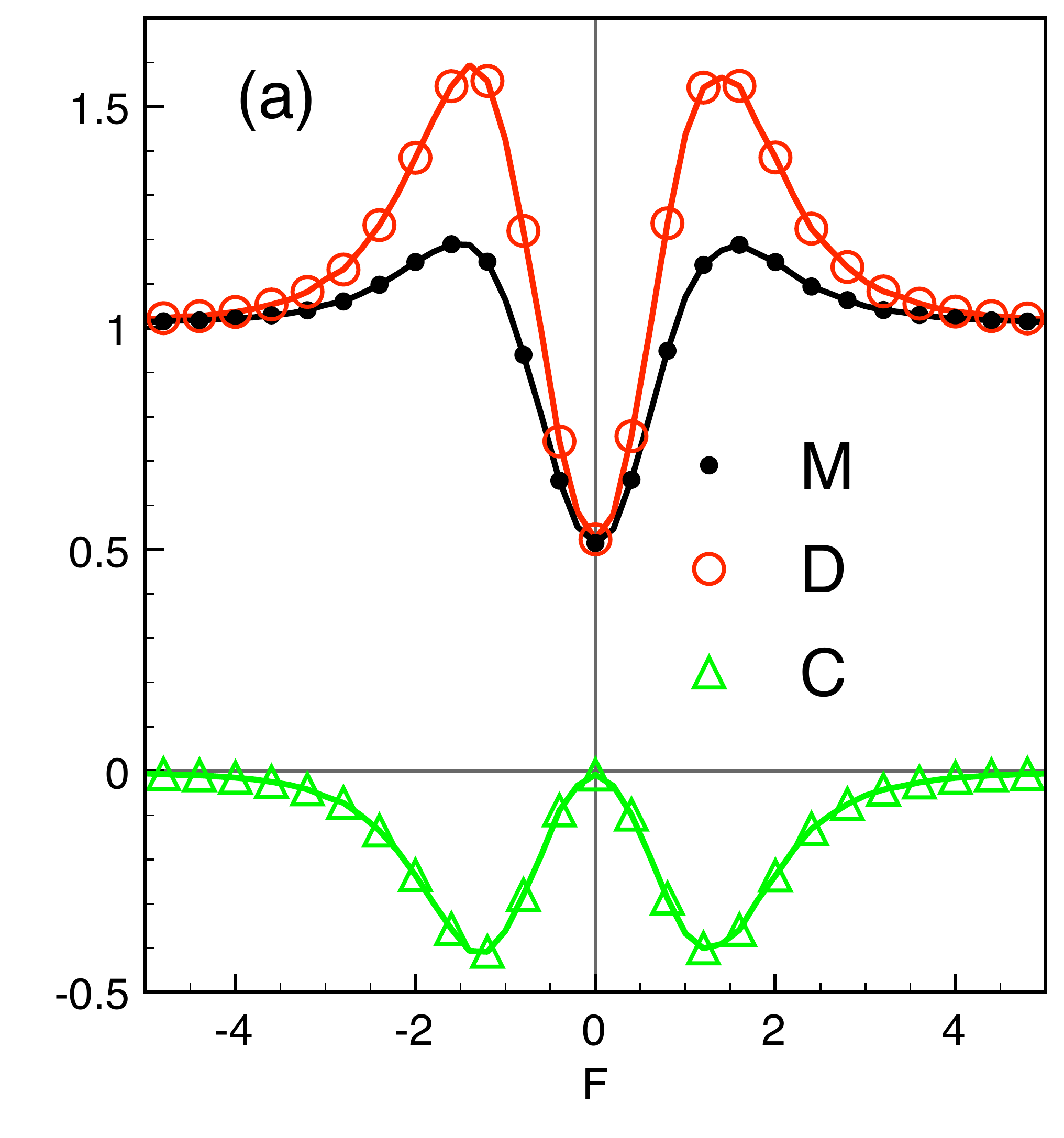}
\includegraphics[width=0.45\textwidth]{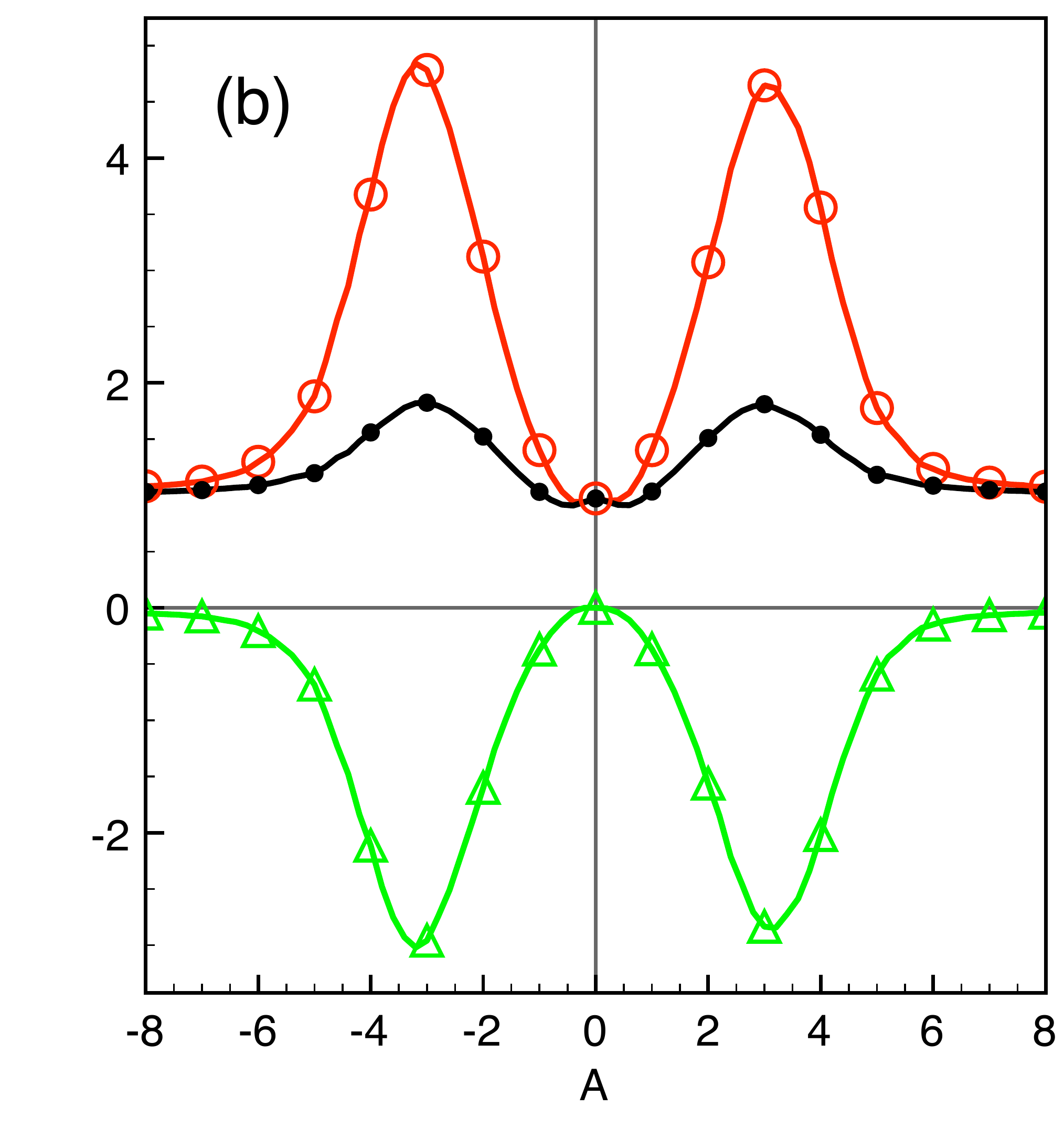}
\caption{The long time limits of the mobility and diffusion are shown. In (a): simulations of a diffusion with a potential $-Fx +\cos x$, where $F$ ranges from $-5$ to $5$.  In (b) the potential is $-A(x-\cos x)$, with $A$ ranging
from $-8$ to $8$. In both cases $T=\gamma = m = 1$. (Online version in color.)}
\label{fig:withforcebis}
\end{figure}

It is clear that the Sutherland--Einstein relation $M=D/T$ is violated. Even though the correction term $C1$ tends
to zero for large times, the correction term $C2$ does not.
However, the correction to the Sutherland--Einstein
relation is not always equally big. E.g. for small forces the correction is small.
For large driving as in Fig.~\ref{fig:withforce} (b), the correction is also small. The force is then so big, that
the potential $U$ does not play an important role. Without the potential $U$ the dynamics of the system is
completely translation invariant, and with a simple change of coordinates one can prove that the Sutherland--Einstein relation
holds. Therefore one expects that in the limit $F\to\infty$ the mobility and diffusion are proportional,
and have the same values as those for pure diffusion.
This can indeed be seen in Fig.~\ref{fig:withforcebis} (a) for the mobility and diffusion
in a range of forces going from $-5$ to $5$.  For a force that is five times the amplitude of the potential,
the mobility and diffusion are very close to 1, the expected values for a pure diffusion. Finally, as also visible in the figure, both $M(t)$ and $D(t)$ are symmetric under the force flip $F\to -F$.  The resulting curves for mobility and diffusion are very similar to the overdamped case, as e.g. in Figure 3 of \cite{ss1}, where the stationary probability density is exactly known.

Fig.~\ref{fig:withforcebis} (b) shows the mobility and diffusion (long time limits) for a force $A(1+\sin x)$,
with $A$ between $-8$ and $8$. Again, for large amplitude $A$ the relation approches that of pure diffusion just as for $F\to\infty$ in Fig.~\ref{fig:withforcebis}.  The reason for that is unclear; the forcing is large but also the potential is large now.  At any rate, the diffusion is clearly much more sensitive to the strength of the force than is the mobility.\\

\begin{figure}[!h]
\centering
\includegraphics[width=0.5\textwidth]{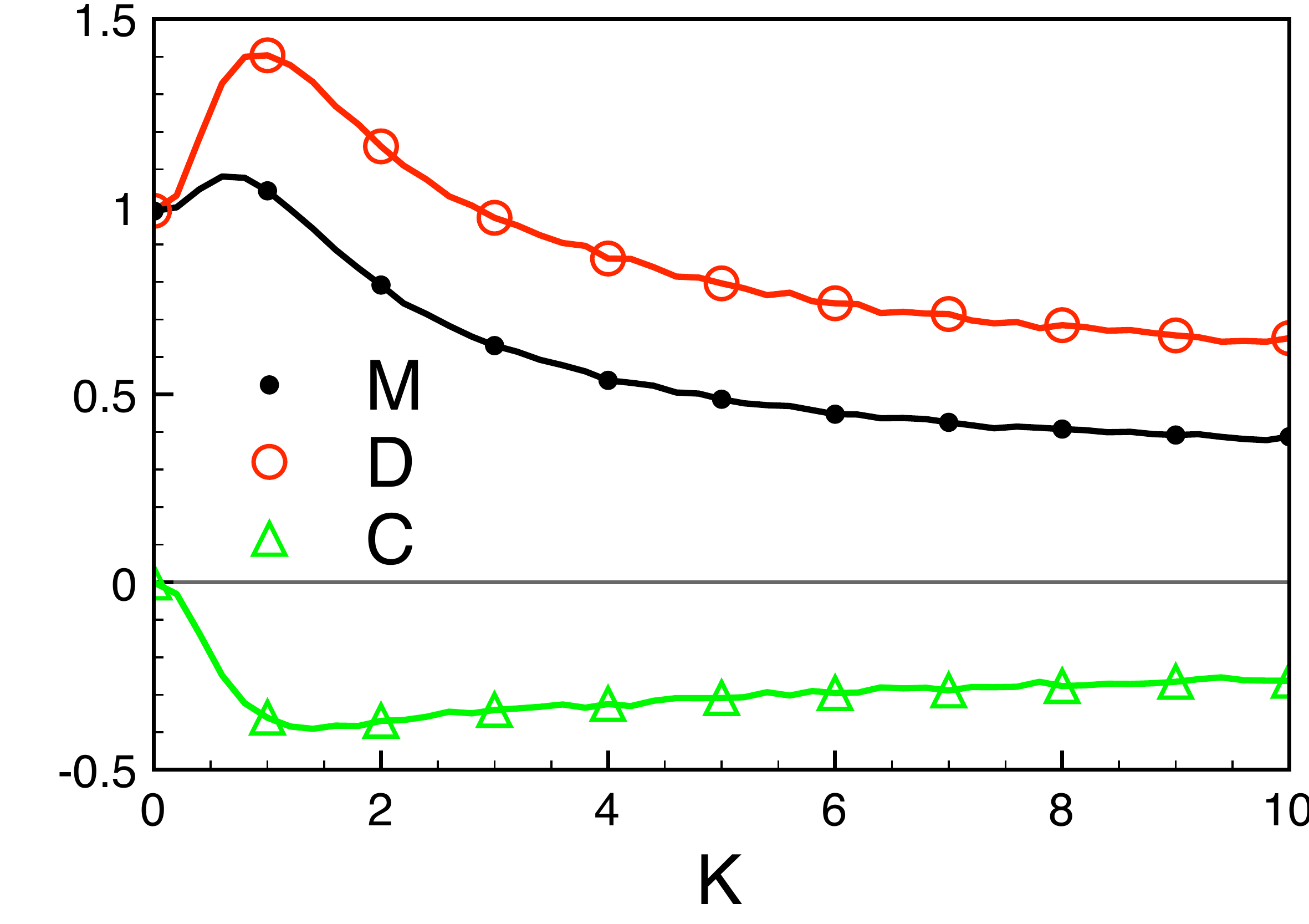}
\caption{The long time limits of the mobility and diffusion are shown for a potential given by $-Fx + \cos Kx$, with
$K$ ranging from $0$ to $10$. Again, we took $T=\gamma = m = 1$. (Online version in color.)}
\label{fig:withforcefreq}
\end{figure}

Finally, Fig. \ref{fig:withforcefreq} shows the mobility and diffusion (long time limits) for a potential
$-Fx + \cos Kx$,
with $K$ between $0$ and $10$. For $K$ between $1$ and $10$ the difference between mobility and diffusion is almost constant.

\subsection{Rotational force}

\begin{figure}[!h]
\centering
\includegraphics[width=0.463\textwidth]{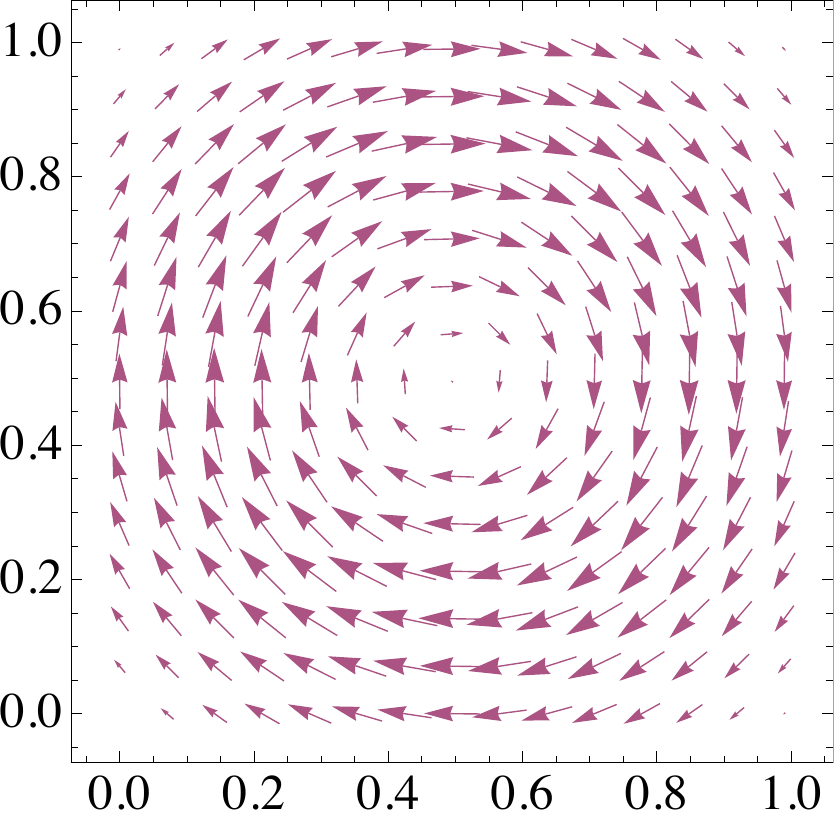}
\hspace{0.5cm}
\includegraphics[width=0.437\textwidth]{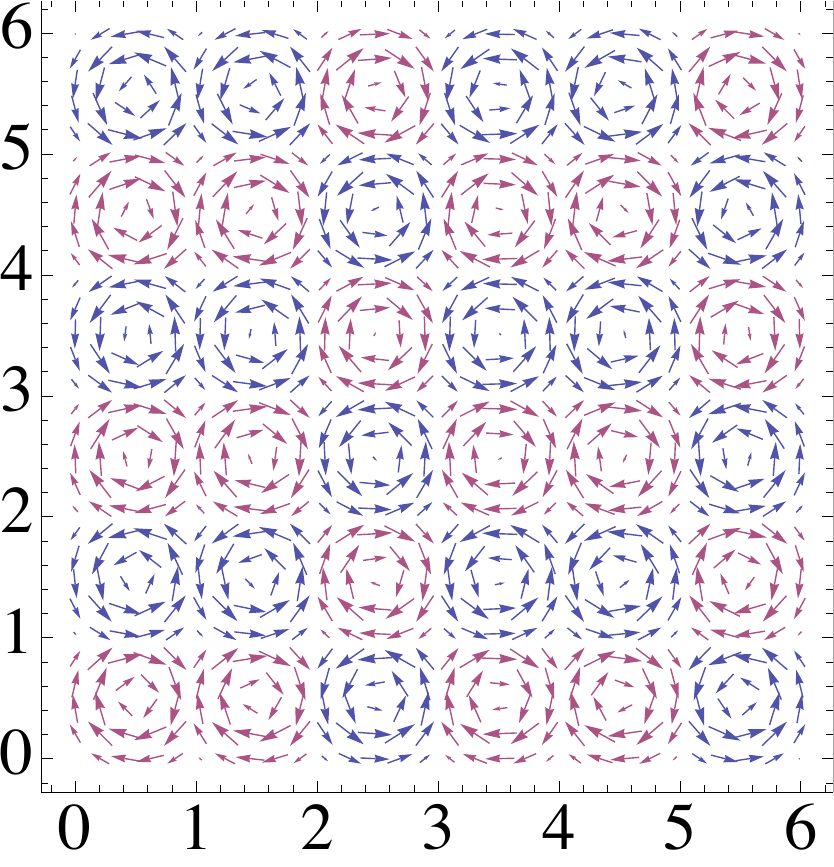}
\caption{Vector plot of the force $F$ in (\ref{curlf}). (Online version in color.)}
\label{fig:forcerotor}
\end{figure}

Apart from forces coming from nonperiodic potentials, nonequilibrium can also be installed from forces with a nonzero curl on the full space.  They induce vortices and let the particles undertake rotational motion, but with more drastic departures from the Sutherland--Einstein relation than in the case (\ref{langmag}) of a magnetic field.\\
As an example we simulated the two-dimensional dynamics
\begin{eqnarray}
m\dot{v}_{x,t} &=& F_x(x_t,y_t) - \gamma v_{x,t} + \sqrt{2m\gamma T}\,\xi_{x,t}\label{curlf}\\
m\dot{v}_{y,t} &=& F_y(x_t,y_t) - \gamma v_{y,t} + \sqrt{2m\gamma T}\,\xi_{y,t}\nonumber
\end{eqnarray}
 The force $\vec{F} = A \vec{f}$, with $A$ some constant and with, for $0 \leq x,y < 1$,
\begin{eqnarray*}
f_x(x,y) &=& (r-\sqrt{2})\, (\frac{1}{2}-y)\\
f_y(x,y) &=& (r-\sqrt{2})\, (x-\frac{1}{2})
\end{eqnarray*}
with $r = \sqrt{(x-\frac{1}{2})^2 + (y-\frac{1}{2})^2}$. This is shown in the left part of Fig.~\ref{fig:forcerotor}.

\begin{figure}[!h]
\centering
\includegraphics[width=0.45\textwidth]{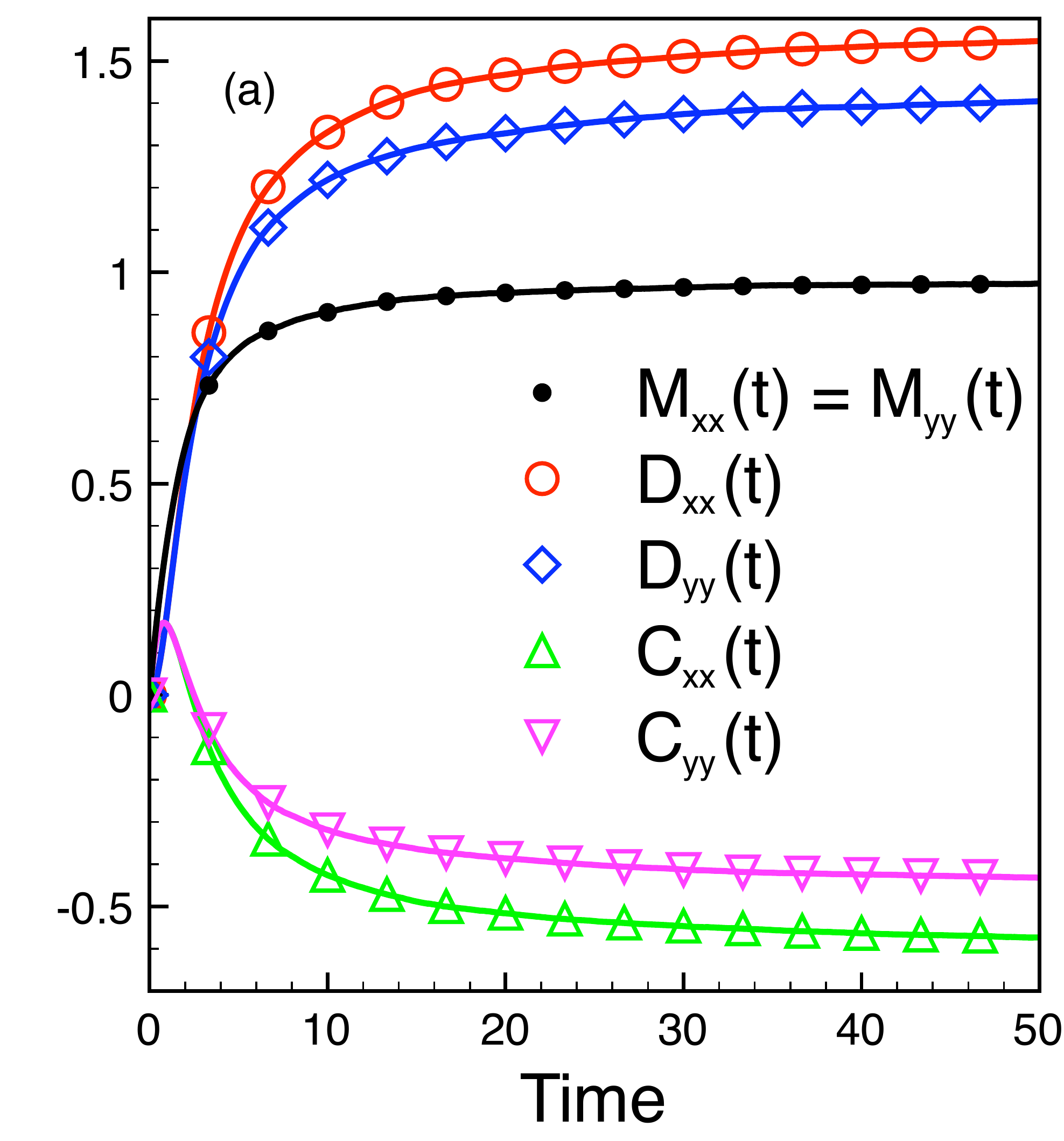}
\includegraphics[width=0.45\textwidth]{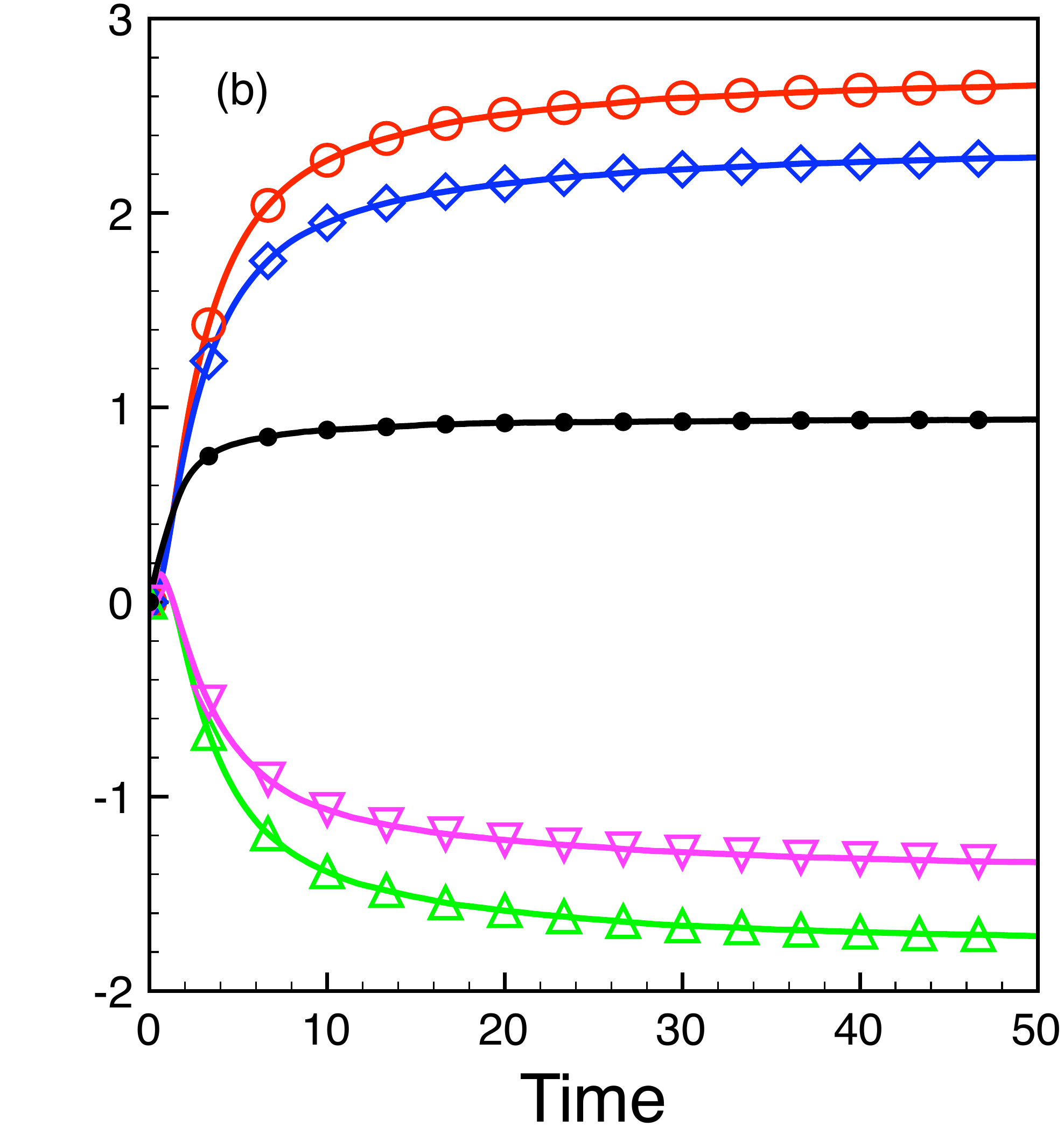}
\caption{Values $T=\gamma=m=1$.
In (a) the constant $A$ is $20$, while in (b) $A=40$. Initial conditions in both cases
take position and velocity equal to zero. (Online version in color.)}
\label{fig:forcerotorsim}
\end{figure}

This force field is repeated outside that unit square, however sometimes in the reverse direction. More precisely,
\begin{eqnarray*}
f_x(x,y) &=& a\,(r-\sqrt{2})\,(y-\frac{1}{2})\\
f_y(x,y) &=& a\,(r-\sqrt{2})\,(\frac{1}{2}-x)\\
a &=& (1-2\delta_{2,x\textrm{\footnotesize{ mod }} 3})\,(1-2\delta_{1,y\textrm{\footnotesize{ mod }} 2})
\end{eqnarray*}
This gives the pattern shown in the right part of Fig.~\ref{fig:forcerotor}. The specific choices are probably not so important. Simulation results for this system are shown in Fig.~\ref{fig:forcerotorsim}.

\begin{figure}[!h]
\centering
\includegraphics[width=0.9\textwidth]{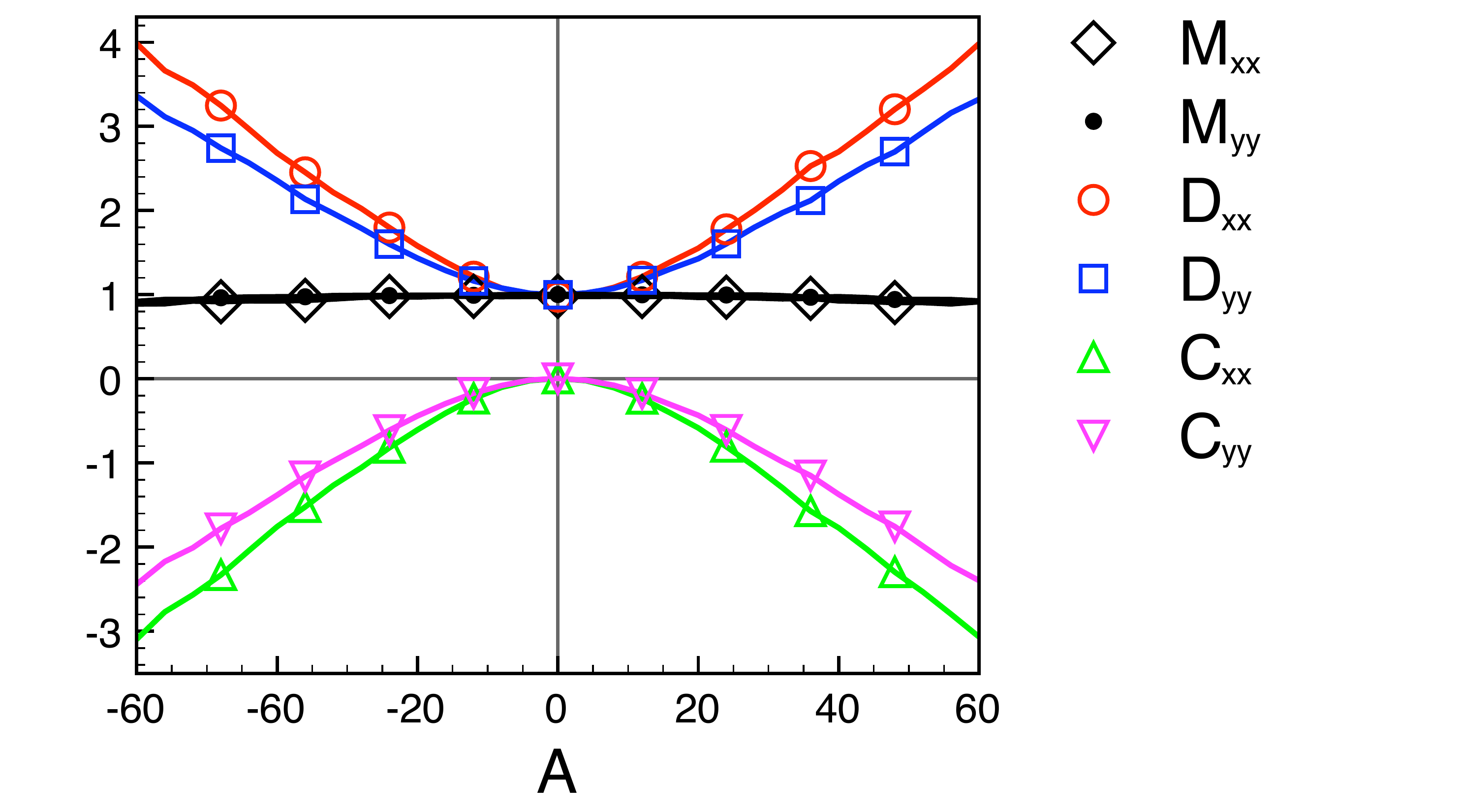}
\caption{Again $T=\gamma=m=1$,
and $A$ ranges from $-60$ to $+60$. (Online version in color.)}
\label{fig:forcerotorf}
\end{figure}

We only show the diagonal elements
of the mobility, diffusion and the correction (C); the off-diagonal elements turn out to be zero.
In both cases the diffusion in the $x$-direction is bigger than in the $y$-direction. Furthermore, when comparing $A=20$ with $A=40$, we see that the diffusion very much depends on the strength of the force, while the mobility
is in both cases approximately the same. To see this more clearly, Fig.~\ref{fig:forcerotorf} shows the large time limits of mobility, diffusion, and their difference, for $A$ in the range $[-60,+60]$.

Observe that the diffusion increases rapidly with the amplitude $|A|$ of the force, while the mobility
remains almost constant (it even decreases a little). In particular, large forcing $A\uparrow \infty$ does not at all lead to pure diffusion here, because the situation is very different from that in Fig.~\ref{fig:withforcebis}.  We also see that all quantities are symmetric under $A \to -A$.
Indeed, reversing the force is the same as translating the whole system, which has no effect on the long time behavior.

\section{The symmetrized mobility}\label{app2}

We now consider the symmetric part of the mobility matrix, i.e., we rewrite equation (\ref{genresult1}) to:
\[\frac{M_{ij}(t)+M_{ji}(t)}{2} = \frac{1}{T}D_{ij}(t) +\frac{1}{4\gamma mT t}\Big[\Big<({\vec r}_t - {\vec r}_0)_i;{\vec \Psi}_j\Big>+\Big<({\vec r}_t - {\vec r}_0)_j;{\vec \Psi}_i\Big>\Big]\]
The last term can be rewritten using elementary algebra:
\begin{eqnarray*}
&&\frac{1}{m\gamma}\Big<({\vec r}_t - {\vec r}_0)_i;{\vec \Psi}_j\Big>+\frac{1}{m\gamma}\Big<({\vec r}_t - {\vec r}_0)_j;{\vec \Psi}_i\Big>\\ && \ \ \ \ \ \ \ \ \ \ \ \ = \Big<\Big[({\vec r}_t - {\vec r}_0)_i + \frac{{\vec \Psi}_i}{m\gamma}\Big];\Big[({\vec r}_t - {\vec r}_0)_j + \frac{{\vec \Psi}_j}{m\gamma}\Big]\Big>\\
&& \ \ \ \ \ \ \ \ \ \ \ \ \  \ \ -\Big<({\vec r}_t - {\vec r}_0)_i;({\vec r}_t - {\vec r}_0)_j\Big> - \frac{1}{m^2\gamma^2}\Big< {\vec \Psi}_i;{\vec \Psi}_j \Big>
\end{eqnarray*}
The second term on the right-hand side is proportional to the diffusion function.
Furthermore, the first of the terms on the right-hand side of the last equation can be simplified: for that, let us write the Langevin equation in its integral form:
\[ \vec{v}_t - \vec{v}_0 = -\gamma (\vec{r}_t-\vec{r}_0) + \int_0^tds\Big[\frac{\vec{F}(\vec{r}_s,\vec{v}_s)}{ m}+\sqrt{\frac{2 \gamma T}{ m}}\vec{\xi}_s\Big] \]
Using definition  (\ref{psi}) of the functional $\Psi$, we substitute
\[ \frac{1}{m\gamma}\vec{\Psi} +  \vec{r}_t-\vec{r}_0 = \sqrt{\frac{2 T}{m\gamma}}\int_0^tds\ \vec{\xi}_s\]
The right-hand side involves Gaussian white noise and $\Big<\xi_{u,i}\xi_{s,j}\Big> = \delta(s-u)\delta_{i,j}$. The result is the following expression for the symmetric part of the mobility:
\begin{eqnarray}\label{resultrewritten}
\frac{M_{ij}(t)+M_{ji}(t)}{2} = \frac{D_{ij}(t)}{2T} + \frac{\delta_{i,j}}{2\gamma m} - \frac{1}{4\gamma^2 m^2T t}\Big< \Psi_i;\Psi_j \Big>
\end{eqnarray}
Using exponential relaxation in large time limits (see below in Appendix \ref{app1}) one can see that in the long time limit this relation becomes
(\ref{symr}).  However, even for finite times when $i=j$, the third term on the right-hand side in (\ref{resultrewritten}) is minus a variance, yielding the following bound for the mobility (diagonal elements):
\begin{equation}\label{bou}
M_{ii}(t) \leq \frac{D_{ii}(t)}{2T} + \frac{1}{2m\gamma}
\end{equation}
with equality for pure diffusion.

\section{Conclusions}
This work investigates the relation between diffusion and mobility for Langevin particles.  The particles are independent and passive, undergoing white noise and friction from the fluid in thermal equilibrium.  The total force on the particles is periodic but not compatible with a periodic potential, bringing the system out-of-equilibrium.  We have studied the modified Sutherland--Einstein relation.  The new mobility--diffusion relation remains explicit and we have visualized the typical dependencies on the nonequilibrium driving and other parameters.  For the future we hope that the inverse analysis will also be possible, i.e., that our formul{\ae} will enable to obtain useful information about unknown aspects of diffusive nonequilibria exactly by measuring the correction to the Sutherland--Einstein relation and from comparing it with (\ref{genresult2}) or with (\ref{symr}).  In many cases however we expect that further extensions to nonMarkovian evolutions will be necessary to meet the physics of small systems immersed in viscoelastic media.

\appendix

\section{Smoothness and mixing}\label{app1}

We argue here why the diffusion matrix $D_{ij}(t)$ has a finite limit and why certain terms do not contribute in the long time limit;
that is the vanishing of (\ref{corrnul}) and (\ref{corrpotzero}).
We do not want to go into full technical details but it should be clear that our results depend on good exponential mixing properties of the dynamics
with propagation of smooth densities.  Of course, there is no stationary regime in unbounded diffusive systems, but this problem is irrelevant whenever expectations are considered of quantities that have the same periodicity as the dynamics (forces) of the system.
(This gives no restriction on the velocity dependence of these quantities.) In that case we can restrict the dynamics
of the system to one period, with periodic boundary conditions (reducing the infinite space to a torus).
Taken on a torus, our system does have a stationary regime, i.e., the Fokker-Planck equation (\ref{fokplagen}) with periodic boundary conditions can in principle be solved with left-hand side zero. The solution is assumed to be the smooth density $\rho(\vec{r},\vec{v})$, giving the stationary distribution of positions (on the torus) and velocities. By assumption then, any function $g$ with the same periodicity as the dynamics satisfies
\[ \lim_{t\to\infty}\left<g(\vec{r}_t,\vec{v}_t)\right> = \int d\vec{r}\, d\vec{v}\;\rho(\vec{r},\vec{v})\,g(\vec{r},\vec{v}) \equiv \left<g\right>_{\rho} \]
and there exist finite constants $C$ and $\alpha > 0$ such that for large enough times $t\geq 0$,
\begin{eqnarray*}
\left|\Big<g(\vec{r}_t,\vec{v}_t)\Big> - \Big<g\Big>_{\rho} \right| &\leq & C\sqrt{\Big<g^2\Big>_{\rho}}e^{-\alpha t}
\end{eqnarray*}
From the above bound, one can deduce bounds on truncated correlation functions.  For example, writing $g(t) = g(\vec{r}_t,\vec{v}_t)$,
for $t\geq s$,
\begin{eqnarray*}
\Big<f(s);g(t)\Big> = \Big<f(s)\Big[\Big<g(t)\Big>_{s}-\Big<g\Big>_{\rho}\Big]\Big> + \Big<f(s)\Big>\Big[\Big<g\Big>_{\rho} - \Big<g(t)\Big> \Big]
\end{eqnarray*}
where $\Big<g(t)\Big>_{s}$ stands for the conditional expectation of $g$ at time $t$, given the state at time $s$.  We can then insert the exponential bounds:
\begin{eqnarray}
\Big<f(s);g(t)\Big> &\leq& \Big<|f(s)|\Big>C\sqrt{\Big<g^2\Big>_{\rho}}\left[e^{-\alpha(t-s)}+ e^{-\alpha t}\right]\nonumber\\
&\leq & \left[ \Big<|f(s)|\Big>_\rho + C'\sqrt{\left<f^2\right>_{\rho}}e^{-\alpha' s} \right]C\sqrt{\Big<g^2\Big>_{\rho}}\left[e^{-\alpha(t-s)}+ e^{-\alpha t}\right]\nonumber\\
&\leq & \sqrt{\Big<f^2\Big>_{\rho}\Big<g^2\Big>_{\rho}}\left[ 1 + C'e^{-\alpha' s} \right]C\left[e^{-\alpha(t-s)}+ e^{-\alpha t}\right]\label{expcorr}
\end{eqnarray}
With this bound, one obtains that the diffusion matrix has a finite limit:
\begin{eqnarray*}
\lim_{t\to\infty}\left|D_{ij}(t)\right| &\leq & \lim_{t\to\infty}\frac{1}{2t}\int_0^tds\int_0^sdu\left[\left|\Big<v_{i,s}v_{j,u}\Big>\right| + \left|\Big<v_{i,u}v_{j,s}\Big>\right|\right]\\
&\leq & \sqrt{\Big<v_i^2\Big>_{\rho}\Big<v_j^2\Big>_{\rho}}\frac{C'\alpha + C\alpha'}{2\alpha\alpha'}
\end{eqnarray*}
  Similarly, for (\ref{corrnul}), we find that
\begin{eqnarray*}
&&\left|\frac{1}{t}\Big<({\vec r}_t - {\vec r}_0)_i;(\vec{v}_t - \vec{v}_0)_j\Big>\right|\\
&& \ \ \ \leq \lim_{t\to\infty}\frac{1}{2t}\int_0^tdu\left[\left|\Big<v_{i,u}v_{j,t}\Big>\right| + \left|\Big<v_{i,u}v_{j,0}\Big>\right|\right]
\end{eqnarray*}
goes to zero, by direct use of (\ref{expcorr}).\\

The proof of equation  (\ref{corrpotzero}) is similar to the one for (\ref{corrnul}), with one extra element,
namely the fact that the dynamics in this case is an equilibrium dynamics, meaning that the associated stationary distribution $\rho$ is the equilibrium distribution. Equilibrium is characterized by time-reversal invariance. This means in particular that the expectation value of a time-antisymmetric quantity is zero in equilibrium. In our case, this means that
\[\Big<({\vec r}_t-{\vec r}_0)_i; \int_0^t ds\nabla_{r_j} U(\vec{r})\Big>_{\rho} = 0 \]
The rest of the proof of (\ref{corrpotzero}) is quite straightforward.

\section{The mobility for one-dimensional diffusions}\label{mo1}
We give a short proof of formula \eqref{fesgal}. Consider the stationary probability current $j^f_{\rho}$ for one-dimensional overdamped diffusion on the circle in the presence of a constant force $f$:
\[ j^f_{\rho} = \chi[f-U']\rho - \chi\,T\,\rho' \]
where $\rho$ is the stationary distribution.  Because of stationarity,
the current does not depend on the spatial coordinate $x$. We divide the last equation by $\rho$ and integrate over the circle:
\[ j^f_{\rho}\int \frac{dx}{\rho(x)} = R\chi f \]
Up to first order in $f$, the stationary current is thus given by
\[ j^f_{\rho} = \frac{R\chi f}{\int \frac{dx}{\rho_0(x)}} \]
where $\rho_0 = e^{-U(x)/T} / \int e^{-U(x)/T}$ is the equilibrium distribution.
This gives the mobility
\[ M =  R\left.\frac{\partial}{\partial f}j^f_{\rho}\right|_{f=0} = \frac{R^2\chi}{\int e^{+U(x)/T}\int e^{-U(x)/T}} \]
because
\[ M = \lim_{t\to\infty} \frac{1}{t}\left.\frac{\partial}{\partial f}\left<x_t-x_0\right>^f\right|_{f=0} = \lim_{t\to\infty} \frac{1}{t}\left.\frac{\partial}{\partial f}\int_0^t\left<dx_t\right>^f\right|_{f=0} \]

\section{About the simulations}

All graphs in this paper are the result of simulations made with the programming language C++ by directly
applying the Langevin equation. This means (taking the simple example of one dimension) taking variables $x$ and $v$ which at each timestep of length $dt$
change by the operations
\begin{eqnarray*}
x &\to & x + vdt\\
v &\to & v + F(x,v)dt -\gamma v dt + \sqrt{2m\gamma T dt}\xi
\end{eqnarray*}
where $\xi$ is a random number drawn with a standard normal distribution (mean zero and variance 1).
In this way a trajectory of consecutive positions and velocities is generated.
In the same way trajectories are generated for a dynamics where a small constant force is added.
The relevant quantities (diffusion, integrated forces, displacement of the particle) are then computed
at each timestep by averaging over many simulated trajectories. The size of the timestep, of the force,
and the number of simulated trajectories varied between the different examples in this text to find each time
a good compromise between statistical accuracy and computation time. The length of the trajectories was taken
such that the quantities showed a clear convergence to a constant value.
As an example, for the case of the nonperiodic potential, we took
$dt = 0.002$, $f = 0.03$, the number of timesteps $15000$ and the number of trajectories $160000$.
As the goal of this paper is not providing accurate quantitative information, but rather a qualitative visualization of the relation between mobility and diffusion, we did not compute the
statistical and systematical errors. Rather, we check each time that the mobility equals the sum of the terms on the right-hand side of (\ref{genresult1}). The mobility needs much more computation time (number of simulated trajectories) than the other quantities. The problem there is that one subtracts two quantities
that are close to each other, and then divides by a small number. Furthermore, for the mobility one needs to simulate two systems: the perturbed and the unperturbed one. This also means that our results are numerically useful, as one does not need to compute the mobility separately. We used this fact in the simulations where the force was varied, as in Figures \ref{fig:withforcebis}, \ref{fig:withforcefreq}, \ref{fig:forcerotorf}. The simulations there took a longer time, as each different value of the force required a different simulation. By leaving out the mobility, the number of trajectories could however be reduced (e.g. for the nonperiodic force the number was $8000$).

\label{lastpage}

\end{document}